\documentclass[twocolumn]{aa}
\usepackage{graphicx}
\usepackage{txfonts}
\usepackage{longtable}
\usepackage{natbib}
\begin{document}
   \title{The extended structure of the remote cluster B514 in M31}

   \subtitle{Detection of extra-tidal stars\thanks{Based on observations with
   the NASA/ESA {\em Hubble Space Telescope} obtained at the Space Telescope
   Science Institute (STScI), which is operated by AURA, Inc., under NASA
   contract NAS 5-26555.}\fnmsep\thanks{Table 5 is only available in electronic
   form at the CDS via anonymous ftp to cdsarc.u-strasbg.fr(130.79.128.5) or via
   http://cdsweb.u-strasbg.fr/cgi-bin/qcat?J/A+A/}}

   \author{L. Federici\inst{1},
          M. Bellazzini\inst{1},
          S. Galleti\inst{1},
          F. Fusi Pecci\inst{1},
          A. Buzzoni\inst{1}
	  G. Parmeggiani\inst{1}}

   \offprints{L. Federici}

   \institute{INAF - Osservatorio Astronomico di Bologna, Via Ranzani 1,
        40127 Bologna, Italy\\
              \email{luciana.federici@oabo.inaf.it}}
   \authorrunning{L. Federici et al.}
   \titlerunning{The extended structure of the cluster B514}
   \date{Submitted 18 Jan 2007; Accepted 6 Jun 2007 }

 
  \abstract
  {}  
  {We present a
  study of the density profile of the remote M31 globular  cluster B514,
  obtained from HST/ACS observations.} 
  {Coupling
  the analysis of the distribution of the integrated light with  star counts we
  are able to reliably follow the profile of the cluster out to $r\sim
  35\arcsec$, corresponding to $\simeq 130$ pc.  The profile is well fitted,
  out to $\sim 15$ core radii, by a King Model having C=1.65.  With  an
  estimated core radius $r_c=0.38\arcsec$, this corresponds to a tidal radius
  of $r_t\sim 17\arcsec$ ($\sim 65$ pc). The analysis of the light profile
  allows also the estimate of the ellipticity and position angle of the
  isophotes within $r\le 20\arcsec$.} 
  {We find
  that both the light and the star counts profiles show a departure from the
  best fit King model for $r\ga 8\arcsec$ - as a surface brightness excess  at
  large radii, and the star counts profile shows a clear break in
  correspondence of the estimated tidal radius. Both features are interpreted
  as the signature of the presence of extra tidal stars around the cluster. It
  is also shown that B514 has a half-light radius significantly larger than 
  ordinary globular clusters of the same luminosity. In the $M_V$ vs. log$r_h$
  plane, B514 lies in a region inhabited by peculiar clusters, like $\omega$
  Cen, G1, NGC2419 and others, as well as by the nuclei of dwarf elliptical
  galaxies.}
   {}


   \maketitle
%

\section{Introduction}

Until a couple of years ago we lack any knowledge of M31 clusters at large
(projected) distances from the center of the galaxy; the farthest known cluster
was G1, at $R_p\sim 35$ kpc from the center of M31, while (a few) Galactic 
globulars are
found out to galactocentric distance of $R_{GC}\sim 120$ kpc (10 at $R_{GC}>30$
kpc, according to Harris \cite{harris}). Recent searches have identified several
new remote clusters\footnote{In the present context we will 
call {\em remote} clusters those lying at galactocentric distances larger 
than $30$ kpc.} in M31 (Huxor et al. \cite{huxout},  Galleti et al. \cite{b514}, 
Bates et al. \cite{bates},  Martin et al. \cite{martin}). These studies seem to
indicate that a relatively large number of globulars are still to be discovered
in the extreme outskirts of M31. A significant sample of distant globulars may
provide extremely useful information on the early evolution of the halo of M31.
In particular, it has been suggested that bright globular clusters may be the
remnants of disrupted nucleated dwarf galaxies (Freeman \& Bland-Hawthorn
\cite{araa}, and references therein; see also Brodie \& Strader \cite{brodie}): 
if this were the case the probability of
finding the observational fingerprints of these kind of phenomena is much
higher at large distances from the center of parent galaxies, where
substructures may survive for long times (Bullock \& Johnston \cite{bullock})
and the overall stellar density is very low.  
Moreover, the structure and evolution of clusters orbiting in very-low density 
environments is a very interesting topic in itself. 

In Galleti et al. (\cite{b514hst}, hereafter G06b) we have presented deep 
Hubble Space Telescope - Advanced Camera for Survey photometry of the recently 
discovered cluster B514 
(Galleti et al. \cite{b514}, hereafter G05), lying at $R_p\simeq 55$ kpc from 
the center of M31. The derived Color Magnitude Diagram (CMD) revealed that B514
is a genuine old and metal poor globular cluster ($[Fe/H]=-1.8$, confirmed also
by the spectroscopic estimate by G05). The cluster is very bright ($M_V\simeq
-9.1$) and appears quite extended, similarly to the brightest remote cluster
of the Milky Way, i.e. NGC~2419. Here we present the analysis of the surface
brightness distribution of B514 obtained from the same HST-ACS data. Coupling
the surface brightness profile obtained from the integrated light - for the 
inner parts - to star counts in the outer region, and thanks to the
extremely low level of background density in the field,
we were able to identify
an unequivocal break in the outer profile of the cluster, indicating the
presence of extra-tidal stars (see Johnston et al. \cite{katy}; Combes, Leon \&
Meylan \cite{combes}; Grillmair et al. \cite{grill,grillm31}; 
Leon, Meylan \& Combes
\cite{leon}). 
Extra-tidal components and/or extended tidal tails have been
observed in several Galactic globulars (see Grillmair et al. \cite{grill};
Leon et al. \cite{leon}; Testa et al. \cite{vicem92}; 
Odenkirchen et al. \cite{pal5}; Lee et al. \cite{lee7492};
Belokurov et al. \cite{belok_a}; and references therein). Holland et al.
\cite{holland} and Barmby, Holland \& Huchra \cite{barmby},
hereafter BHH, found some M31 clusters whose
light profile exceeds the best fit King \cite{king62,king} 
model in the outermost
regions, and interpreted this discrepancy as an extra-tidal component. 
Grillmair et al. \cite{grillm31} found the same kind of discrepancy in three M31
clusters; they were able to follow the density profile of the clusters to
significantly below the background level by coupling the light profile with the
profile obtained from star-counts. 
By applying a similar technique, we are able to follow 
the profile of B514 out to $r\sim 35\arcsec$.
Moreover, we found that B514 has a half-light radius
($r_h$) larger (by $\ga 15$\%) than typical globular clusters
of the same luminosity, a characteristic shared by a few very peculiar systems,
like $\omega$ Cen, M54, G1 and NGC~2419 (see Mackey \& van den Bergh
\cite{macksyd}, hereafter MB05, and Hasegan et al. \cite{hase}, 
for a thorough discussion).

\section{The surface brightness profile of B514}

Since the details of the observations and data reduction are reported in 
G06b, here we recall just a few essential elements, referring the interested
reader to that paper.
The cluster has been observed with the Wide Field Channel of the ACS. The WFC
has a total field of view of $202\arcsec \times 202\arcsec$ and a pixel scale of
$0.049\arcsec$ pixel$^{-1}$.
The observational material is constituted by three F606W images (total
$t_{exp}=2412$ s) and three F814W images (total $t_{exp}=2419$ s), and the
associated combined (drizzled) images. The cluster is placed in the center of
one of the two ACS/WFC chips (Chip 2), while Chip 1 sample the field
population (see G06b and below). 
If not otherwise stated, magnitudes are always in the VEGAMAG scale
as defined by Sirianni et al. \cite{sirianni}. The reddening corrections are
performed as in G06b, assuming $E(B-V)=0.10$; a distance modulus $(m-M)_0=24.47$
is also assumed, after McConnachie et al. \cite{mccon}, corresponding
to $D_{\sun}=783$ kpc. At this distance $1\arcsec$ correspond to 3.8 pc and one
ACS/WFC pixel to 0.19 pc.

The extreme crowding conditions prevent the full resolution into stars of the
densest region of M31 globular clusters, even at the exquisite spatial 
resolution achieved by HST cameras. However, very accurate and well
resolved  surface brightness profiles can be obtained studying the distribution
of their integrated light (see, for example, Fusi Pecci et al. \cite{strufoc}; 
Djorgovski et al. \cite{djorM31}; BHH).
This technique, very successful in the bright inner regions of the clusters, is
limited in the outermost part of the clusters, where the low luminosity density
coming from cluster stars may be overwhelmed by the brightness of the
background. In this regime star counts may be much more efficient, since cluster
stars may be easily identified out to large radii, under favorable conditions
(see Grillmair et al. \cite{grillm31}).
The ACS field studied here lies more than three degrees 
apart from the center of M31 and it appears to have an exceedingly low 
density of background stars (see G06b and below). 
This allowed us to derive a reliable light profile,
completely unaffected by incompleteness, out to $r\sim 20\arcsec$, and to extend
the analysis out to $r\sim 35\arcsec$, in completely uncrowded regions, by
counting stars having colors and magnitudes typical of the cluster population.

\subsection{The light profile}

The light profile has been obtained independently from the F606W and F814W
drizzled images.
A few heavily saturated foreground
stars (all at $r>26\arcsec$ from the cluster center) have been excised from the
images and replaced with the mean value of the surrounding background, to avoid
contamination of the profile. 
 
\begin{figure}
   \centering
   \includegraphics[width=\columnwidth]{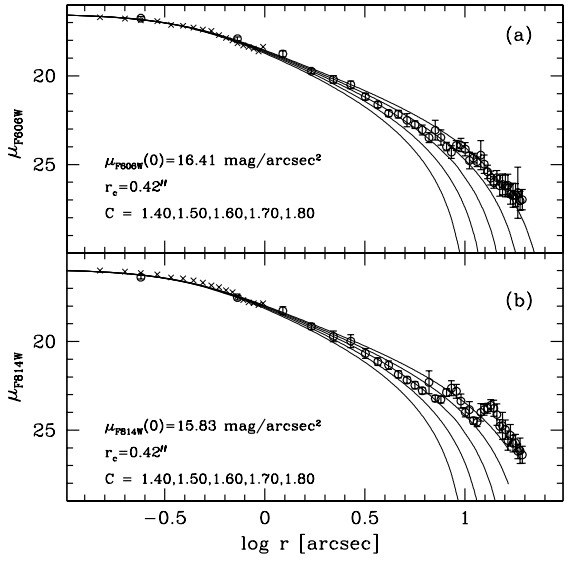}
      \caption{Surface brightness profiles (in mag/arcsec$^2$) of B514
      in F606W (upper panel) and F814W (lower panel). The "x"s in the
      innermost region are from the 1-pixel step profile, while the open circles
      are average values over 10 px bins. The curves are 
      PSF-convolved King models
      of different concentration, from C=1.4 to C=1.7, from left to right.
      The adopted core radius and central surface brightness are also reported.
              }
         \label{proflight}
   \end{figure}

The light profiles were derived using the XVISTA software, maintained by 
J. Holtzman\footnote{See {\tt http://astronomy.nmsu.edu/holtz/xvista/index.html}
for download and documentation}. XVISTA iteratively resolves the profile by
fitting ellipses to the observed light distribution until a stable solution is
reached (see Fusi Pecci et al. \cite{strufoc}, Djorgovski et al. \cite{djorM31},
for examples of applications to M31 clusters, and Lauer \cite{lauer}, 
for a detailed description of the usage). The code provides as output, the
coordinates of the center, the surface brightness, the ellipticity
($\epsilon=1.-b/a$), and the position angle (PA, in degrees, measured
anti-clockwise from the North direction) of each fitted ellipse, as well as
the total amount of light enclosed within each ellipse. The profiles are derived
with a single pixel step. This resolution is appropriate for the innermost
regions of the cluster ($r\sol 2\arcsec-3\arcsec$) where the light intensity 
is very
high, while at larger radii provides a quite noisy profile. This problem is
solved by getting an average of the whole profile over 10 px bins: this sampling
ensure a satisfying level of noise out to $r\sim 20\arcsec$. The background
level is estimated as the average surface brightness in large ($\sim 100\times
100 ~px^2$) ``empty'' areas far away from the cluster. The level of background
is very low ($\mu_{F606W}\sim 29.5$ mag/arcsec$^2$) and the final
background-subtracted profiles were verified to be very robust to variations of
the adopted background. The uncertainty on the position of the center is of
order of $\sim 2$ px in the X and Y directions: variations on the assumed
position of this size doesn't affect significantly the derived profiles.
The derived intensity profiles can be directly converted into 
magnitude/arcsec$^{-2}$ units using the VEGAMAG zero points of Sirianni et al.
\cite{sirianni}. The derived F660W and F814W profiles are shown in 
Fig.~\ref{proflight}. The "x"s in the innermost $2\arcsec$ are from the 
original 1-pixel step profiles, while the open circles with error bars are the
averages in 10 pixels bins. The profiles are quite smooth, well-behaved and
very similar in shape, at least out to $r\simeq 8\arcsec$, i.e. the radius
enclosing $\simeq 90$\% of the whole cluster light. Outside this radius both
profiles show a marginal excess of light with respect to the best-fitting 
King \cite{king62} model (see below), but the F814W profile appears more noisy,
probably because of the larger weight associated to the contribution of
individual bright RGB stars in this redder passband. 
{\bf The F606W light profile is reported in Tab.~3.} The apparent integrated
magnitudes in the VEGAMAG system $mag_{t,F606W}$ and $mag_{t,F814W}$ were estimated by 
integrating the respective light profiles.

%

We take the Half Width at Half Maximum (HWHM) of the profile as the core radius (King
\cite{king62}, Spitzer \cite{spitzer}). Once fixed this parameter we searched
for the King's models providing the best-fit to both profiles.
In order to take in the proper account the ACS Point Spread Function (PSF) in
measuring the light profile, the King models have been convolved with 
analytic F606W/F814W
PSFs models, modeled on observed bright stars, as done by 
Barmby et al.\cite{bar07}.
All the comparisons between observed profiles and King's models presented in the
following involve only PSF-convolved theoretical profiles. In Tab.~1 we report
both the {\em observed} and the {\em de-convolved} best-fit 
parameters (see below). The
former must be adopted when dealing with the observed profiles, while the latter
must be used in the comparisons with other clusters. We note that the adoption
of PSF-convolved profiles results in small changes ($\la$ 10\%) in the cluster
parameters, as usually occurs for extended M31 clusters like B514 (see Barmby 
et al.\cite{bar07}).

The derived F660W and F814W profiles are shown in 
Fig.~\ref{proflight}, compared to PSF-convolved King's models 
with concentration parameter $C=log(r_t/r_c)=1.4,1.5,1.6,1.7$. If we 
limit to the most reliable region within
$r=8\arcsec$ (encompassing $\simeq 19$ core radii), the best fit is achieved
with the $C=1.7$ model for the F606W profile and $C=1.6$ for 
the F814W profile.

We can gain some insight of the
uncertainty associated with many observed/derived parameters by the comparison
between the estimates obtained in the F606W and F814W profiles. For size
parameters, (core radius $r_c$, half-light radius $r_h$, tidal radius $r_t$, see
King \cite{king62}) we adopt the mean of the independent estimates obtained by 
the two profiles. For example, for the core radius we obtain $r_c=0.38\arcsec$
from F606W and $r_c=0.45\arcsec$ from F814W, and we adopt 
$r_c=0.42\pm0.03\arcsec$; adopting C=1.65
we obtain a tidal radius $r_t\simeq 18.8\pm 2.5\arcsec$;
for the half-light radius we obtain $r_h=1.52\arcsec$ 
from F606W and $r_h=1.73\arcsec$ from F814W, and we adopt 
$r_h=1.6\pm 0.2\arcsec$ (observed values, see Tab.~1). 

The half-light radii have been
computed also by performing aperture photometry on circular concentric annuli;
this independent procedure provided the same results obtained with XVISTA,
indicating that the estimate of this parameter is very robust.
Also the estimates of the apparent integrated magnitudes have been checked in
this way, and the results obtained with different methods are fully consistent.

A summary of the measured structural parameters is presented in Tab.~1, while
Tab.~2 shows the derived parameters, i.e. those involving assumptions on distance
and reddening and/or transformations to the standard Johnson-Kron-Cousins 
photometric system. The latter are achieved with the transformations presented
in G06a. Note that if Sirianni et al.'s transformations are used instead,
slightly brighter V magnitudes are obtained (by $\sim 0.06$ mag), while the
final I magnitudes are the same to within $\pm 0.02$ mag. 
$r_c$, $r_t$ and surface brightness measures reported in Tab.~2 are derived from
average de-convolved values. $\mu_{r_h}$ is the mean surface brightness within
$r_h$, while $\mu(0)$ is the central value of the surface brightness.

In a very recent paper, Mackey et al \cite{mackhux1} reported on the CMDs and 
(observed) half-light radii and integrated magnitudes of eight M31 clusters, 
including B514 (their GC4), from  independent  ACS observations. 
The parameters of B514 obtained by this team 
are in good agreement with those presented by us here and in 
Galleti et al.\cite{b514hst}.

\subsubsection{Color profile, ellipticity and position angle}

\begin{figure}
   \centering
   \includegraphics[width=\columnwidth]{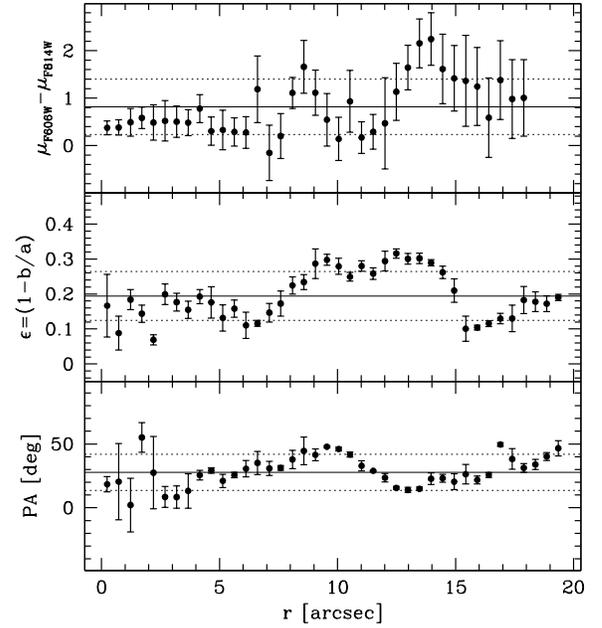}
      \caption{Upper panel: color profile of B514; 
      the continuous line is the overall 
      mean color and the dotted lines encloses the $\pm 1$ standard deviation
      range. Middle panel: ellipticity profile. Lower panel: position angle
      profile. The latter two profiles are the average of the profiles obtained
      from the F606W and F814W images. The meaning of the lines is the same as
      in the upper panel.
              }
         \label{color}
   \end{figure}

In the upper panel of Fig.~\ref{color} we show the color profile of B514.
The profile is remarkably constant in the inner $6\arcsec$, while it becomes
very noisy in the outer regions. It is clear that in the low-surface-brightness
outer parts the contribution of individual RGB/BHB stars may be important to
establish the local color. A general tendency toward redder colors in the outer
regions can be noted, a phenomenon that has been observed also in other
clusters (see Djorgovski et al. \cite{dgradient}, Djorgovski \& 
Piotto \cite{gradient}, and Barmby et al. \cite{barmby}).
However, in the present case, we regard this trend as of marginal significance,
given the uncertainties. 

%

The middle and lower panels of Fig.~\ref{color} display the ellipticity and
position angle profiles, respectively. In both cases, once verified that the
profiles obtained from the F606W and F814W were fully consistent, we averaged
the two. B514 has a mean ellipticity $\langle \epsilon \rangle = 0.19\pm 0.07$,
quite high for a globular cluster, but not extraordinary (MB05).
A sizable enhancement of the ellipticity is apparent 
between $r\sim 8\arcsec$ and $r\sim 15\arcsec$. A twist of the isophotes seems 
to occur in the same radial range (lower panel, PA changing by $\sim 30\degr$),
suggesting a disturbed morphology in this range. The overall conclusion is that
the cluster is rather elongated in the NW-SE direction, as is apparent by 
simply looking at the image (G06b).

%
\begin{table*}[!ht]
\caption{Observed and de-convolved parameters}
\label{tab1}
\centering
\renewcommand{\footnoterule}{}  
\begin{tabular}{lr@{.}lr@{.}lr@{$\pm$}lr@{.}l}
\hline \hline
         &\multicolumn{6}{c}{ ---------------~~Observed~~--------------}&\multicolumn{2}{c}{De-convolved} \\
Parameter&\multicolumn{2}{c}{\phantom{55}F606W}&\multicolumn{2}{c}{\phantom{14.}F814W}&\multicolumn{2}{c}{Average} &\multicolumn{2}{c}{Average}  \\
\hline
$r_c$ [arcsec]      &\phantom{154}0  & 38 &\phantom{147}0  &45 &\phantom{1}0.42&0.03& \phantom{1555}0&38\\
$r_h$ [arcsec]      &\phantom{154}1  & 52 &\phantom{145}1  &73 &\phantom{1}1.6 &0.2 & \phantom{1555}1&44\\
$r_t$ [arcsec]      &\phantom{154}19 & 0 &\phantom{145}18 &0 &18.8&2.5& \phantom{155}17&0\\
C          &\phantom{154}1  &  7	  &\phantom{147}1  &  6 	&\multicolumn{2}{c}{~~~1.65}& \phantom{1555}1&65\\
$\mu(0)$ [mag/arcsec$^{2}$]   &\phantom{154}16 & 41	  &\phantom{147}15 &83  &\multicolumn{2}{c}{  }&\multicolumn{2}{c}{16.33/15.74 $^{\mathrm{a}}$}\\
$\mu_{r_h}$ [mag/arcsec$^{2}$] &\phantom{154}18 & 5	  &\phantom{147}17 &9  &\multicolumn{2}{c}{  }&\multicolumn{2}{c}{18.4/17.5 $^{\mathrm{b}}$}\\
$mag_{t}\ [VEGAMAG]$     & \multicolumn{2}{r}{15.48$\pm$0.06} & \multicolumn{2}{r}{14.71$\pm$0.06}  &\multicolumn{2}{c}{  }&\multicolumn{2}{c}{  }\\
\end{tabular}
\begin{list}{}{}
\item[$^{\mathrm{a}}$]  De-convolved F606W/F814W central surface brightnesses. 
\item[$^{\mathrm{b}}$]  De-convolved F606W/F814W half-light radius surface brightnesses.
\end{list}
\end{table*}
%

%
\begin{table}
\begin{minipage}[t]{\columnwidth}
\caption{Derived Parameters}
\label{tab2}
\centering
\renewcommand{\footnoterule}{}  
\begin{tabular}{lrl}
\hline \hline
Parameter&\multicolumn{2}{r}{Average/Adopted}\\
\hline
$M_V$      &\multicolumn{2}{c}{$-9.1\pm0.1$~~}\\
$M_I$      &\multicolumn{2}{c}{$-9.9\pm0.1$~~}\\
$\mu_V(0)$ [mag/arcsec$^{2}$] &\multicolumn{2}{c}{\phantom{1}16.5 $^{\mathrm{a}}$}   \\
$\mu_I(0)$ [mag/arcsec$^{2}$] &\multicolumn{2}{c}{\phantom{1}15.7 $^{\mathrm{a}}$ } \\
$\mu_{V,r_h}$ [mag/arcsec$^{2}$] &\multicolumn{2}{c}{\phantom{1}18.4 $^{\mathrm{a}}$}  \\
$\mu_{I,r_h}$ [mag/arcsec$^{2}$] &\multicolumn{2}{c}{\phantom{1}17.5 $^{\mathrm{a}}$} \\
$r_c$ [pc]      &\multicolumn{2}{c}{\phantom{1}1.4$^{\mathrm{a}}$}\\
$r_h$ [pc]      &\multicolumn{2}{c}{\phantom{1}5.4$^{\mathrm{a}}$} \\
$\langle$Ellipticity$\rangle$ &\multicolumn{2}{c}{\phantom{5}$0.19\pm0.07$~}\\
$\langle$Position Angle$\rangle$ [deg]&\multicolumn{2}{c}{\phantom{1}$28\pm14$~} \\
\hline
\end{tabular}
\end{minipage}
\begin{list}{}{}
\item[$^{\mathrm{a}}$]  De-convolved quantities. 
\end{list}
\end{table}
%

\subsection{The profile from star counts}

\begin{figure}
   \centering
   \includegraphics[width=\columnwidth]{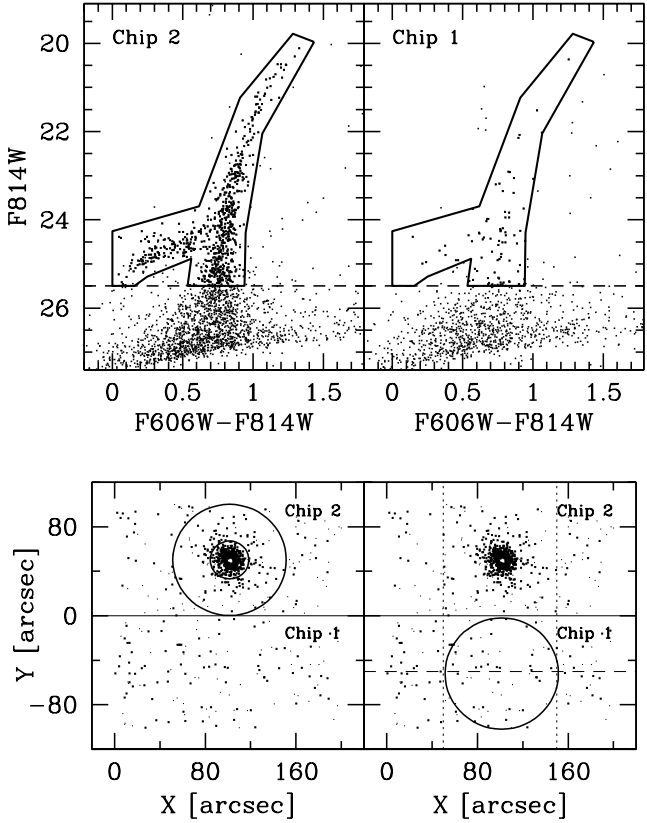}
      \caption{Selection of likely cluster members on the CMD. Upper panels:
      CMDs of the stars in the ACS/WFC chip containing the cluster (left panel)
      and of those in the ``empty'' chip (right panel). The contour is the
      filter adopted to select star for further analysis; the dashed line marks
      the adopted magnitude limit (F814W$<25.5$). Lower panels: maps of the
      selected (heavy dots) and unselected (but brighter than  F814W$=25.5$)
      stars of Chip 2 (above the continuous horizontal line) and  Chip 1 (below
      the continuous horizontal line).  Here we adopted a representation that
      approximately displays the actual relative position of the two chips. 
      The regions selected for the analysis are enclosed within the large
      circles (radius of $50\arcsec$). This is the largest circle that can be
      enclosed in a single chip. The smaller circle in the lower-left panel has
      $r=20\arcsec$ and is indicative of the derived tidal limit of the
      cluster. The dotted vertical lines and the horizontal dashed line plotted
      in the lower right panel show the portions of the field that have been
      used to obtain estimates of the background level, that were compared to 
      that obtained in the  $50\arcsec$ circle in Chip 1 (see text).
              }
         \label{sel}
   \end{figure}

For star counts we adopt the same catalog as G06b, including only well 
measured stars (see G06b, for details). Table 5 (online material) reports
the photometry of the individual stars. In the upper panels of Fig.~\ref{sel}
we show the CMDs for the chip containing the cluster (Chip 2) and for the chip
presumably sampling the field population (Chip 1, see G06b).  The
reported contour is the filter we adopt to select likely cluster members on the
CMD: it encloses Red Giant Branch (RGB) and Horizontal Branch (HB) stars
having  $F814W\le 25.5$. The filter efficiently excludes obvious color outliers
and  faint stars whose membership can be uncertain. In the lower panels the X,Y
map of the two samples - in their relative positions - is presented.
The horizontal continuous line marks the boundary between the two chips. 
The stars selected by the filter are
plotted as heavy dots. The larger circle has a radius of $50\arcsec$ and is the
largest circle that can be fully enclosed within one WFC chip. The following
analysis is restricted only to filter-selected stars within this circle. The
background level of the stellar density is estimated from the whole $50\arcsec$
circle in Chip 1 {\bf as $\rho_{bkg}=0.0043\pm 0.0013$ stars/arcsec$^2$}, 
while we derive the cluster profile from selected stars in
Chip 2. It can't be excluded that cluster stars are present also in Chip 1.
However, estimating the background level in the $X<50\arcsec$, $X>150\arcsec$
regions of both chips (enclosed by dotted lines in the lower right panel of
Fig.~\ref{sel}), and in the $Y<-50\arcsec$, $Y>-50\arcsec$ regions of Chip 1
(separated by the long dashed horizontal line in the lower right panel 
of Fig.~\ref{sel}), we found that the background is the same as that measured in
the $50\arcsec$ circle, within $< 2-\sigma$, {\bf ranging from 
0.0024$\pm 0.0010$ stars/arcsec$^2$ to 0.0056$\pm 0.0015$ stars/arcsec$^2$},
and there is no discernible density
gradient outside $r=50\arcsec$ from the cluster center. To have a more reliable
estimate of the background level we would need observations of a larger (or more
distant) field that, unfortunately, is lacking. However, this implies that, if
anything, we are slightly {\em overestimating} the background. The
possible associated bias would act {\em against} the detection of feeble
extra-tidal components, hence it cannot be at the origin of the excess 
of surface brightness at large radii that is discussed below. 

%

\begin{figure}
   \centering
   \includegraphics[width=\columnwidth]{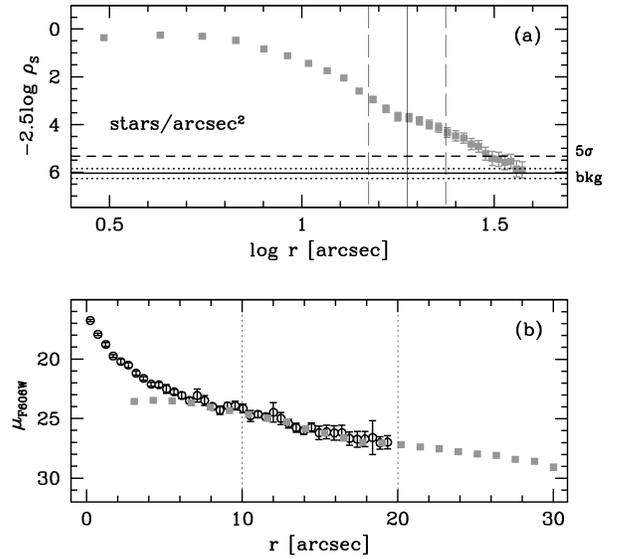}
      \caption{Panel (a): observed surface density profile obtained from star
      counts (grey squares). 
      The continuous horizontal line marks the background level, the
      dotted lines marks the $\pm 1-\sigma$ levels around the background, and the
      dashed line marks the $5-\sigma$ level  above the background. The
      continuous vertical line marks the position of the tidal radius as derived
      from the fit of the light profile, assuming C=1.6, the long-dashed lines
      show the position of $r_t$ under the assumptions C=1.55 (left) and C=1.75
      (right).
      Panel (b): Comparison and match between the F606W light profile (empty
      circles) and the (background-subtracted) star counts profile (grey squares). The scale of the
      horizontal axis has been kept linear to provide a clearer view of the
      superposition between the two profiles. The dotted lines encloses the
      region in which the match between the two profile is excellent.
              }
         \label{profcounts}
   \end{figure}

The derived profile is shown in  Fig.~\ref{profcounts}a. The profile is very
extended: a level of 5-$\sigma$ above the background is reached at 
$r\simeq 31\arcsec$, while the (adopted) background level is reached at $r\ga
35\arcsec$. This plot shows one of the main results of the present paper: 
not only the observed profile clearly extends much beyond the tidal radius
derived from the light profiles but, above all, {\em a clear change of slope
is detected at $r\sim 18\arcsec$}, i.e. near $r_t$ itself. Note that the excess
between $r\simeq 18\arcsec$ and $r\sim 30\arcsec$ is many $\sigma$ above the 
background, hence it is very significant, even if it encloses just a tiny 
fraction of the total cluster light. 

\begin{figure}
   \centering
   \includegraphics[width=\columnwidth]{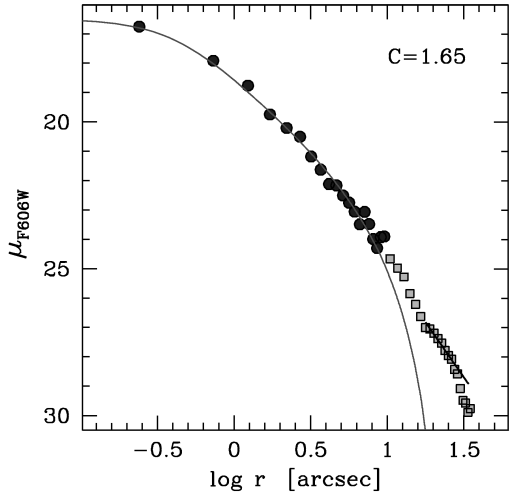}
      \caption{Total surface brightness profile obtained by joining
      the light (dark grey circles) and the star-counts (light grey squares)
      profiles. The light curve is the (PSF-convolved) 
      King model best fitting the light
      profile. The dark segment in the outer part is a 
      {\em surface density} $\propto r^{-3}$ power law. 
              }
         \label{proftot}
   \end{figure}

Given the extreme crowding conditions in the inner part of the
cluster and the strong density gradient, it is expected that the completeness of
the sample is subject to radial variations.  Fig.~\ref{profcounts}b clearly
illustrates the actual case by comparing the  light profile (that is completely
unaffected by incompleteness) and the background-subtracted
 star counts profile {\bf (reported in Tab. ~4)}.
A linear radial scale
is adopted to provide a clearer comparison. This plot shows that incompleteness
significantly affects starcounts for $r\la 6\arcsec$, becoming more and more
important toward the center of the cluster, until, as said, it reaches 100\% at
$r\la 2\arcsec$. However an {\em excellent} match of the profiles can be achieved
in the range $10\arcsec\le r< 20\arcsec$ (i.e. to the end of the light
profile). This clearly proofs that for $r\ge 10\arcsec$ there is no more 
variation
of the incompleteness with radius and, consequently, {\em star counts provide a
fair and fully reliable description of the real profile} in the considered
range. The vertical shift applied to the star counts profile to match the light
profile automatically provides also the normalization constant to transform
surface stellar densities into surface brightness. 
Therefore, the two profiles can
be joined into one, covering the full $0\le r \le 35\arcsec$ range, as shown 
in Fig.~\ref{proftot}. 

%
%

%
\begin{table}
\begin{minipage}[t]{\columnwidth}
\caption{Background-subtracted F606W Surface Brightness profile from the integrated light.}
\label{tab3}
\centering
\renewcommand{\footnoterule}{}  
\begin{tabular}{ccc}
\hline \hline
   r      &$\mu_{F606W}$ & $\epsilon_{\mu}$   \\
 arcsec   &mag/arcsec$^2$&mag/arcsec$^2$ \\
\hline
 0.24& 16.7& 0.1\\
 0.73& 17.9& 0.1\\
 1.23& 18.8& 0.2\\
 1.71& 19.7& 0.1\\
 2.20& 20.2& 0.2\\
 2.69& 20.5& 0.2\\
 3.18& 21.2& 0.2\\
 3.67& 21.6& 0.2\\
 4.16& 22.1& 0.2\\
 4.65& 22.2& 0.2\\
 5.14& 22.5& 0.4\\
 5.63& 22.8& 0.2\\
 6.12& 23.1& 0.3\\
 6.61& 23.5& 0.3\\
 7.11& 23.1& 0.6\\
 7.59& 23.5& 0.4\\
 8.09& 24.0& 0.2\\
 8.57& 24.3& 0.4\\
 9.06& 23.9& 0.3\\
 9.55& 23.9& 0.4\\
10.05& 24.1& 0.4\\
10.53& 24.8& 0.5\\
11.02& 24.6& 0.3\\
11.51& 24.8& 0.3\\
12.01& 24.5& 0.8\\
12.49& 25.0& 0.5\\
12.98& 25.4& 0.4\\
13.47& 25.8& 0.4\\
13.96& 26.0& 0.3\\
14.45& 25.7& 0.4\\
14.94& 26.2& 0.6\\
15.43& 26.1& 0.6\\
15.92& 26.2& 0.6\\
16.41& 26.2& 0.6\\
16.90& 26.7& 0.6\\
17.40& 26.7& 0.7\\
17.89& 26.7& 0.6\\
18.38& 26.6& 1.4\\
18.86& 27.0& 0.5\\
19.35& 27.0& 0.6\\
\hline
\end{tabular}
\end{minipage}
\end{table}

Is it possible that the excess component beyond $r_t$ and/or the observed change
of slope can be due to sources unrelated to the cluster? This possibility is
very hard to conceive, since (1) the adopted quality selections (G06a) and the
CMD filter limit the analysis to relatively bright, well behaved stars that 
shouldn't suffer from any serious contamination, and (2) 
it is very hard to imagine a ``field'' population whose surface density 
decreases with distance from the cluster center. We must conclude that the
detected surface density excess at large radius and the change of slope in 
the profile are genuine properties of the cluster. 
A change of slope in the outer regions of the surface density 
profile is generally interpreted as the signature of the presence of 
tidally stripped stars (see Combes et al. \cite{combes},
Johnston et al. \cite{katy}; Leon et al. \cite{leon}, and references therein).
For brevity, in the following we will refer to the stars beyond the break
in the profile as to extra-tidal stars.
%
\begin{table}
\begin{minipage}[t]{\columnwidth}
\caption{Background-subtracted F606W Surface Brightness profile from star counts.}
\label{tab4}
\centering
\renewcommand{\footnoterule}{}  
\begin{tabular}{ccc}
\hline \hline
   r      &$\mu_{F606W}^{\mathrm{~~~~~~~~~a}}$ & $\epsilon_{\mu}$   \\
 arcsec   &mag/arcsec$^2$&mag/arcsec$^2$ \\
\hline
10.41& 24.7& 0.1\\
11.64& 25.0& 0.1\\
12.86& 25.3& 0.1\\
14.09& 25.8& 0.1\\
15.31& 26.2& 0.1\\
16.54& 26.6& 0.2\\
17.76& 27.0& 0.2\\
18.99& 27.1& 0.2\\
20.21& 27.2& 0.2\\
21.44& 27.4& 0.2\\
22.66& 27.5& 0.2\\
23.89& 27.8& 0.2\\
25.11& 28.0& 0.2\\
26.34& 28.1& 0.2\\
27.56& 28.4& 0.2\\
28.79& 28.6& 0.2\\
30.01& 29.1& 0.3\\
31.24& 29.5& 0.3\\
32.46& 29.6& 0.3\\
33.69& 29.9& 0.3\\
34.91& 29.8& 0.3\\
\hline
\end{tabular}
\begin{list}{}{}
\item[$^{\mathrm{a}}$]  {\bf The star counts profile has been converted into surface brightness 
units (mag/arcsec$^2$)
with the relation $ \mu_{F605W}=-2.5log(\rho_S)+23.20 $, according to the
normalization shown in Fig.~\ref{profcounts}.}
\end{list}
\end{minipage}
\end{table}

\section{Discussion}

A general prediction of incompleteness
theoretical studies of tidal tails is that the slope
of the surface brightness profile is different for bound stars and extra-tidal
stars
(see, for example, Combes et al. \cite{combes}, C99; 
Yim \& Lee \cite{yim}; Johnston et al. \cite{katy}; Montuori et al.
\cite{montuori}). 
Johnston et al. \cite{katy} predicts that the
surface density of stars in the tails should decrease as $r^{-1}$, in
agreement with most observations of extra-tidal stars around Galactic globulars
(Grillmair et al. \cite{grill}; Leon et al. \cite{leon}; Testa et al.
\cite{vicem92}; Odenkirchen et al. \cite{pal5}). On the other hand
C99 conclude that that such shallow slope is to be
expected in the tidal tails at large distances from the parent cluster - 
i.e. fully unbound independent tidal debris -, while in the vicinity of the 
cluster - i.e. immediately beyond the tidal radius - the density should 
decrease as 
$r^{-\alpha}$ with $\alpha=3$ or larger, and the involved stars cannot be
considered as completely unbound. C99 explain the
discrepancy with the observed slopes as due to imperfect subtraction of a very
noisy background, typical in most Galactic cases. It is interesting to note that
in the present case, where we deal with stars in the proximity of the cluster and
the background is virtually non-existing, the density of extra-tidal component
decreases significantly faster than $r^{-1}$, and it is compatible with
$\rho_S\propto r^{-3}$ (the thick segment superposed to the outer profile in
Fig.~\ref{proftot}).

In any case, the only possible alternative to explain the observed excess of
stars and the change of slope in the profile at $r_t\simeq 18\arcsec$ it is to
postulate that the cluster is embedded in a very low surface brightness stellar
system, that is, for instance an unknown dwarf galaxy. We consider this
hypothesis as unlikely, since (a) the density of the detected 
extra-tidal component decreases with distance from the center of the cluster,
that would be possible only if the cluster resides at the center of the
hypothesized system,(b) the surface brightness of the extra-tidal component
is $\ga 26$ mag/arcsec$^2$, significantly lower than the typical central surface
density of local dwarf spheroidals (Mateo \cite{mateo}; but see Zucker 
et al. \cite{zucker}, for a counter-example in M31), and (c) the INT/WFC
survey (Ibata et al. \cite{iba}) reaches the surroundings of B514 (the cluster
is near one edge of the survey field) and no particular substructure is detected
in this region. The twisting and distortion of the outer isophotes 
noted in Sect.~2.1.1, above, also militates for the tidal origin of the observed
structures. 
The detection of tidal tails on opposite sides with respect to the
center of the cluster would have provided a conclusive argument in this sense
(see Montuori et al. \cite{montuori}, and references therein).
Unfortunately the extra-tidal over-density appear as statistically significant
only when azimuth-averaged. We find just a small degree of anisotropy in the
spatial distribution of extra-tidal stars: they appear to be located
preferentially along the SE - NW direction, but the effect is not statistically
significant.

\subsection{Bright clusters and the nuclei of dwarf ellipticals}

Extragalactic surveys performed with high spatial resolution cameras are
revealing the existence of new kinds of stellar systems. The Extended
Clusters (ECs)
found by Huxor \cite{hux} are a typical example, but other kinds of extended
clusters have been identified in more distant galaxies (see Peng et al.
\cite{peng}, for a thorough review and discussion).
Hilker et al. \cite{hilker} and
Drinkwater et al. \cite{drink} recently discovered a new kind of dwarf galaxies 
inhabiting galaxy clusters that are slightly more luminous than the brightest
globular clusters and significantly more compact than any dwarf galaxy, the
Ultra Compact Dwarf galaxies (UCD). 
It has been suggested that such systems are the
dense remnant of tidally harassed nucleated galaxies (Drinkwater at al.
\cite{ucd}). In this context a new perspective on the possible relations and
differences among these various system is gradually emerging, and the comparison
among the structural properties may provide interesting insights in this sense.

   \begin{figure}
   \centering
   \includegraphics[width=\columnwidth]{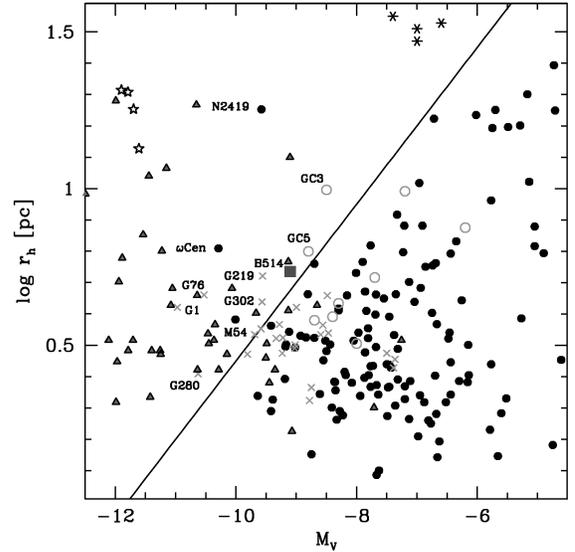}
      \caption{Absolute integrated V magnitudes vs. logarithm of the half light
      radius in pc for various low-luminosity stellar systems. The plotted line
      is the threshold for ordinary globular clusters in this plane as defined
      by MB05,
      log $r_h$ = 0.25$M_V$+2.95.
      Filled circles are Galactic globular clusters, from MB05;
      grey crosses are M31 globulars,
      our own estimates from HST images; asterisks are ECs from Mackey et al.
      \cite{mackhux}; open circles are outer M31 GCs from Mackey et al.
      \cite{mackhux1};
      open stars are UCDs in the Fornax cluster from 
      De Propris et al. \cite{depropris}; red triangles are nuclei of
      dwarf elliptical galaxies in the Virgo cluster from C\^ot\'e et al.
      \cite{cote}; the empty circles are the M31 clusters of 
      Mackey et al.~\cite{mackhux1};the grey square is B514. 
      The globular clusters lying above
      the line are labeled with their names.
              }
         \label{mvrh}
   \end{figure}
%

In this line, MB05 
compared globular clusters
from different galaxies in the $M_V$ vs. $logr_h$ diagnostic plane (see also
Hasegan et al. \cite{hase}). 
They found
that the bulk of globulars lies below the line log$r_h$=0.25$M_V$+2.95. All the
objects that are found above this threshold, namely $\omega$ Cen, NGC~2419, M54
and G1, are very bright and anomalous clusters: all of them were previously
indicated as possible remnants of disrupted nucleated dwarf galaxies (Freeman \&
Bland-Hawthorn \cite{araa}. The
$M_V$ vs. $log r_h$ plane has a distinct advantage with respect to other similar
diagnostic planes (as, for instance, $M_V$ vs $\mu_V(0)$, Kormendy \cite{kor}),
since $r_h$ is a quite easy-to-measure and reddening independent quantity.

In Fig.~\ref{mvrh} we show the position of B514 and other interesting systems
in the $M_V$ vs. $log r_h$ plane. Filled circles are Galactic globular clusters, 
from MB05;
grey crosses are M31 globulars,
our own estimates from HST images\footnote{These estimates have been obtained
with a profile analysis strictly homogeneous to that performed here for B514;
the results have been compared with the independent estimates by BHH: the
agreement between the observed parameters from the two sources is very good.};
open circles are the M31 clusters of Mackey et al. \cite{mackhux1}, seven of
them having $R_p> 30$ kpc
(in the following we will refer to these clusters as M-GC1, M-GC2, ...,
M-GC10); 
asterisks are ECs from  Mackey et al.
\cite{mackhux}; open stars are UCDs in the Fornax cluster from De Propris et 
al. \cite{depropris}. 
The continuous line is the Mackey \& van den Bergh
threshold mentioned above. Apart of B514, the real novelty of Fig.~\ref{mvrh}
with respect to previous versions (MB05;
Huxor et al. \cite{hux}; Belokurov et al. \cite{belok_b})
is that for the first time it is possible to report also the position of the 
nuclei of dwarf nucleated ellipticals (red triangles). 
C\^ot\'e et al. \cite{cote} measured
half-light radii and {\em g$_{AB}$, z$_{AB}$}\footnote{g$_{AB}$ and z$_{AB}$
are the ACS/WFC F475W, F850LP passbands, respectively, calibrated in the 
ABMAG system; see C\^ot\'e et al. \cite{cote}.}
integrated magnitudes of the 
{\em nuclei}  of several dwarf ellipticals in Virgo from deep ACS/WFC images. 
We converted {\em g$_{AB}$} magnitudes into V magnitudes 
with the transformation:
$$ V=g_{AB}-0.31(g_{AB}-z_{AB}) ~~~~(r.m.s.=0.1~~mag)$$
that we have obtained from 166 bright stars of the cluster NGC2419 that are 
in common between
the g$_{AB}$, z$_{AB}$ photometry we obtained from archive ACS/WFC data and the
secondary standards provided by Stetson \cite{stet} for this cluster.
Then, we converted the integrated V
magnitudes into absolute magnitudes by adopting the reddening and distance 
modulus provided by C\^ot\'e et al. \cite{cote}.  

There are a number of very interesting indications emerging from
Fig.~\ref{mvrh}:

\begin{enumerate}

\item{} In addition to the clusters already noted 
        by MB05,
        i.e. M54, $\omega$ Cen, NGC~2419 and G1,
        there are a few other bright M31 globulars lying above the ``
	ordinary globular cluster'' threshold: B514, 
	M-GC3, M-GC5, G76, G280, G219, and G302.
	All of these systems are classified as globular clusters but have
	half-light radii significantly larger than those of typical {\em
	genuine} globular clusters of the same metallicity.

\item{} The distribution of nuclei nicely overlaps the position of these
        anomalous clusters. In particular, the nuclei of dwarf ellipticals
	appear to join the brightest globulars to the anomalous
	``above-threshold'' clusters and to the UCD galaxies, which have been
	also interpreted as the nuclear remnants of shredded galaxies
	(Drinkwater et al. \cite{ucd}). This is the first time that a clear
	connection between the structure of nuclei, UCDs and anomalous clusters
	is directly established by comparing sizable samples.
	It would be of great interest to extend the comparison to velocity
	dispersions, but unfortunately these quantities are not available for
	the sample of nuclei considered here (but see Hasegan et al. \cite{hase})
	and for most of M31 globulars, including B514 .
	
\item{}	ECs, on the other hand, seem to have a different nature,
        somehow intermediate between globular and open star clusters
	(Peng et al. \cite{peng}). 	
	
\item{} The remote clusters B514 ($R_p\simeq 55$ kpc) and M-GC5
       ($R_p\simeq 78$ kpc) lie above the log$r_h$ - $M_V$ threshold, while
       M-GC1 ($R_p\simeq 46$ kpc) and M-GC10 ($R_p\simeq 100$ kpc) are located
       well below the threshold, fully immersed within the distribution of 
       ordinary globular clusters. This demonstrates that a large half-light
       radius is not a distinctive characteristic of remote clusters.
	
\item{}	It is interesting to note that the faint
	Galactic satellites recently discovered by various SDSS teams (see
	Belokurov et al. \cite{belok_b}, and references therein) lie outside
	the limits of Fig.~\ref{mvrh}, all of them having log$r_h>$ 1.6 and
	$M_V>-8$, just above ECs. Ordinary dwarf spheroidals have similar sizes
	but they have $M_V<-8$ (Mateo \cite{mateo}).

\end{enumerate}

It is interesting to have a closer look to the anomalous ``above-threshold'' 
clusters.
$\omega$ Cen and G1 clearly host stars of different chemical composition
(and, presumably, age), hence - unlike classical globulars - they were
able to sustain chemical evolution (see, Sollima et al. \cite{sol}; 
Bekki \& Freeman \cite{bekki}; Meylan et al. \cite{g1}, and references 
therein). Both clusters are quite elliptical in shape, similar to B514.
M54 resides within the
nucleus of the Sagittarius dwarf spheroidal (Monaco et al.
\cite{lorenzo} and there are indications of a small metallicity spread
among its stars (Sarajedini \& Layden \cite{sl95}). G76 is a relatively metal
poor cluster ([Fe/H]$\simeq -1.3$, Rich et al. \cite{rich}, hereafter R05) that
is projected onto a very dense star forming region in the disc of M31
(Bellazzini et al. \cite{micm31}). The extreme crowding conditions in this region
prevents a clear interpretation of the wide RGB shown in its CMD as obtained by
R05, but this feature clearly leaves room for a possible metallicity spread.
G280 is a quite metal rich ([Fe/H]$\simeq -0.5$, R05) and bright cluster; like 
G76 it is projected onto a high density background. G302 is a metal poor cluster
([Fe/H]$\simeq -1.7$, R05) with a blue horizontal branch. Its CMD (R05) is quite
clean and it suggest that G302 is a normal single-population cluster. G219 is a
metal poor cluster ([Fe/H]$\simeq -1.9$, R05) located at $\sim 20$ kpc from the
center of M31. Bellazzini  et al. \cite{micm31} noted that G219 is projected
onto the giant stream discovered by Ibata et al. \cite{iba}. It is remarkable,
in the present context, that both G302 and G219 are among the rare M31 clusters
with a detection of extra-tidal stars, as B514 (Holland et al. \cite{holland}; 
Grillmair et al. \cite{grillm31}). NGC2419 is a remote metal-poor Galactic
globular whose CMD is very similar to that of B514 (Harris et al. \cite{n2419}).
As B514 it show no obvious sign of a metallicity spread, but this may be very
difficult to find out from the CMD of very metal-poor clusters (see MB05).
It has to be noted that NGC2419 has a half-light radius as large as those of the
largest dE nuclei and UCD galaxies.

In summary, B514 fully lies into the region of the $M_V$ vs. log$r_h$ plane that
appear forbidden to ordinary clusters, in company of a few other clusters, many
of which present some kind of peculiarity. The fact that the same region hosts
also the nuclei of dwarf ellipticals provide support to the hypothesis that
these bright extended clusters can be the remnant of disrupted galaxies (Freeman
\& Bland-Hawthorne \cite{araa}; MB05, and references therein; see Brodie \&
Huchra for a thorough discussion on the role of globular clusters within a
cosmological context). 
On the other hand, it is possible that ordinary clusters
are allowed to attain extended structures if their orbit never drives them in
the inner part of the parent galaxy, as it may be the case of NGC2419 and B514
(see van den Bergh, Morbey \& Pazder \cite{vdb}).
However, the coexistence of
ordinary and ``above-threshold'' clusters in the outermost regions of M31 
(see point 4, above) does not support this hypothesis.

In any case the present study confirms that globular clusters in the outskirts
of M31 may reveal many interesting features, thanks also to the favorable
observing conditions (as for instance, the very low background). Remote clusters
are rare in the Milky Way, and they are typically quite faint, with the only
exception of NGC2419 (see Mackey et al. \cite{mackhux1} and Galleti et al.
\cite{g07}). Hence, M31 may provide the opportunity of a systematic study of 
a kind of stellar system that is very rare in the Galaxy. 

Using the same selection criteria that
lead us to the discovery of B514 (G05) we have selected a conspicuous number of
candidate clusters at large distances from the center of M31; at present we have
obtained spectroscopic confirmation (as in G06) for {\em four} of them, all
being nearly as luminous as B514 (Galleti et al. \cite{g07}). 
The follow-up of these remote clusters, as well as of those discovered by 
other teams, may hopefully open a new window in the study of the M31 system 
and of bright clusters as a whole.

\begin{acknowledgements}
We are grateful to Dougal Mackey and Avon Huxor for useful discussions. 
We acknowledge the financial support to this research  by Agenzia Spaziale
Italiana (ASI) and of the Italian Ministro dell'Universit\`a e della Ricerca 
under the grant INAF/PRIN05 1.06.08.03. Part of the data analysis has been
carried out with software developed by P. Montegriffo at INAF -  Bologna
Observatory. 
\end{acknowledgements}

\bibliographystyle{aa}

\end{document}